%% file: main.tex
\documentclass[11pt,twocolumn,journal]{IEEEtran}

\usepackage{algorithm}
\usepackage{algorithmicx}
\usepackage{array}
\usepackage{amsmath}
\usepackage{amssymb}
\usepackage{bm}
\usepackage{citesort}
\usepackage{color}
\usepackage{epsfig}
\usepackage{epstopdf}
\usepackage{graphicx}
\usepackage{multicol}
\usepackage{multirow}
\usepackage{url}
\usepackage[table]{xcolor}
\usepackage{tablefootnote}
\usepackage{threeparttable}

\newlength{\figurewidth}
\newlength{\smallfigurewidth}

\begin{document}

\title
{\huge{Multi-scale PIIFD for Registration of Multi-source Remote Sensing Images}}

\author{
Chenzhong Gao,
Wei Li
\thanks{
C. Gao and W. Li are with the School of Information and Electronics, Beijing Institute Technology, Beijing 100081 China (e-mail: liwei089@ieee.org). }
}

\maketitle
\thispagestyle{empty}
\pagestyle{empty}

\begin{abstract}
\input{abstract}
\end{abstract}


\section{Introduction}
\label{sec:introduction}
\input{Chap.1}

\section{Related Works}
\label{sec:related}
\input{Chap.2}

\section{Proposed Registration Method}
\label{sec:proposed}
\input{Chap.3}

\section{Experimental Results and Discussion}
\label{sec:results}
\input{Chap.4}

\section{Conclusions}
\label{sec:conclusions}
\input{Chap.5}

\bibliographystyle{IEEEtran}
\bibliography{refs}

\end{document}

%% file: abstract.tex
This paper aims at providing multi-source remote sensing images registered in geometric space for image fusion. Focusing on the characteristics and differences of multi-source remote sensing images, a feature-based registration algorithm is implemented. The key technologies include image scale-space for implementing multi-scale properties, Harris corner detection for keypoints extraction, and partial intensity invariant feature descriptor (PIIFD) for keypoints description. Eventually, a multi-scale Harris-PIIFD image registration algorithm framework is proposed. The experimental results of four sets of representative real data show that the algorithm has excellent, stable performance in multi-source remote sensing image registration, and can achieve accurate spatial alignment, which has strong practical application value and certain generalization ability.

%% file: Chap.1.tex
Image registration has always been one of the most challenging works in image processing, especially for multi-source images. In order to achieve in-depth analysis, it is necessary to take advantage of the images obtained by different sensors \cite{zhang2018feature,zhao2020joint,liu2020joint,zhang2020transfer}, and the accurate registration of these images is the fundamental premise. Multi-source image registration is an essential step to implement multi-source data fusion, multi-source collaborative classification, and joint analysis.

In general, satellite remote sensing data provide spatial calibration information (spatial reference), such as latitude and longitude, map grid, etc. But in practice, it is found that there can be problems of spatial information loss. Besides, the spatial reference of different satellites or sensors often has the problem of inconformity, with a difference of a few pixels or even large deviations. It leads to inaccurate ground object correspondence, which can cause a decline in the accuracy of image processing results in data fusion, collaborative classification, change detection, image mosaic, and other applications.

In practical work, remote sensing image registration is usually carried out manually, by using ground control points, or through observation of the human eye, which is time-consuming, laborious, and inefficient. The existing automatic registration algorithm cannot meet the requirements of multi-source remote sensing image registration in accuracy, stability, efficiency, and adaptability. Many mature algorithms are dependent on the single-mode properties of images \cite{lowe2004distinctive,bay2006surf,rublee2011orb,leutenegger2011brisk,alahi2012freak}. However, multi-source remote sensing image registration is a more complex problem in which image pairs are generally obtained from different sensors, times, and viewpoints. Thus, differences between the imaging principle, resolution, condition, and many other factors can cause a multi-source remote sensing image pair to contain changes in scale, blur, rotation, geometric and radiance distortion, content details, and noise, etc, which invalidate the algorithms.

Multi-source remote sensing images are rich in types and have different characteristics, bringing great difficulty to realize universal registration algorithm. For example, hyperspectral image (HSI) and multispectral image (MSI), which provide wealthy spectral information, and light detection and ranging (LiDAR) data, which provide elevation information, are used together for joint classification  \cite{zhang2018feature,zhao2020joint,liu2020joint}. Correspondence analysis with optical and synthetic aperture radar (SAR) images is investigated in depth \cite{zhang2020transfer}. In addition, the fusion of infrared and optical images is also one of the most important and commonly used technical means \cite{ma2019infrared}. Panchromatic(PAN) images with high spatial resolution and texture information are often used for fusion with HSI and MSI. Therefore, it is in great need of effective, universal, and robust algorithms for multi-source remote sensing image registration.

In this paper, focusing on the characteristics and differences of multi-source remote sensing images, a robust feature-based image registration algorithm is implemented, in which we use Harris corner detector for keypoints extraction, scale-space to implement multi-scale properties, and PIIFD for key points description. Eventually, a multi-scale PIIFD (MS-PIIFD) image registration algorithm framework is proposed, which can solve the main problems and difficulties: 1) different scales and blur, 2) radiance distortion, 3) geometric distortion, 4) different content details. The experimental results of 15 scenes of representative data show that the algorithm has excellent, stable performance in multi-source remote sensing image registration, which can achieve accurate spatial alignment, and has good universality and strong practical application value. 

%% file: Chap.2.tex
\begin{figure*}[t]
\centering
\centerline{\includegraphics[width=17cm]{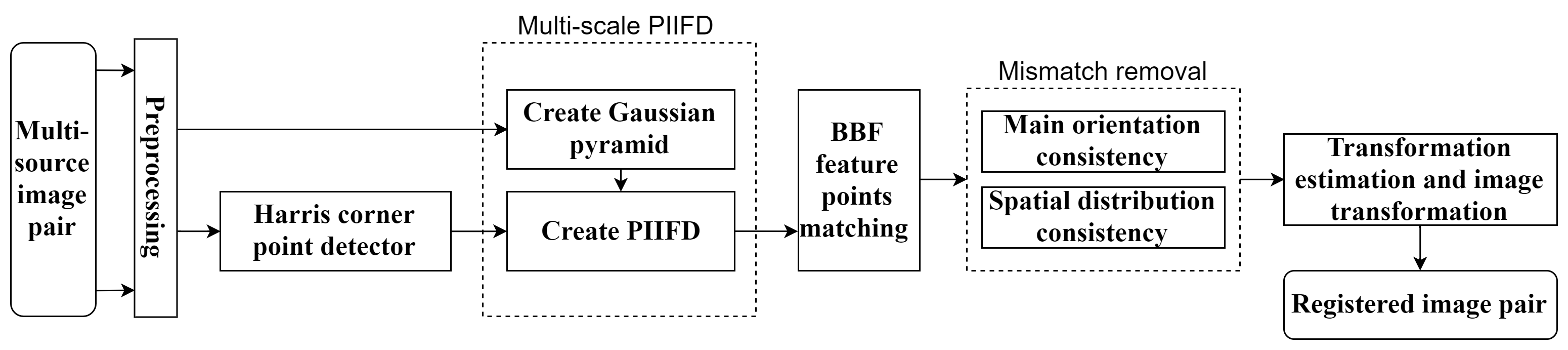}}
\caption{The proposed multi-source remote sensing image registration framework of MS-PIIFD.}
\label{fig:res}
\end{figure*}

Existing automatic image registration algorithms can be generally divided into two categories, the traditional algorithms and the algorithms based on deep learning. Traditional registration algorithms are systematically summarized and classified into area-based methods and feature-based methods according to technical means \cite{zitova2003image}.

\subsection{Area-Based Methods}
\label{ssec:areabased}
The area-based registration methods, also called intensity-based methods, register images by directly using correlation of pixel values, mutual information, or transform domain of the image pairs. Usually, the global image or a predefined local window is used to search for matching \cite{gruen2012development}. By establishing the similarity relationship of the original pixel values \cite{pratt1974correlation,mahmood2011correlation}, statistical properties \cite{viola1997alignment,pluim2003mutual,oliveira2014medical}, or transform domain, such as Fourier transformation \cite{de1987registration,tong2015novel}, and wavelet transformation \cite{le2002automated}, the optimal spatial transformation model between the image pairs is found through parameter optimization methods. The advantage of the area-based methods lies in high registration accuracy, but the drawbacks are particularly obvious. The drawbacks of the area-based methods mainly lie in three aspects.

The first drawback is that this type of method only works for simple spatial changes, such as translation and rotation, which are called rigid transformations. If the image pairs contain more complex spatial changes, such as scale difference, affine or perspective transformation, viewpoint changes, and various nonlinear geographic distortions, it is almost impossible to calculate the transformation parameters by comparing the similarity of the information in a window, which will make the methods fail.

The second drawback, which is the most serious, is that measuring similarity based on pixel values makes such algorithms too dependent on the modality of the images. In other words, area-based methods cannot handle intensity differences. For multi-source images, especially multi-sensor images, because of the different purpose and principle of image collection, their pixel values represent different physical information, so that there are multi-modal properties between them. Take optical and thermal infrared image registration as an example. The former not only has higher spatial resolution and clearer edge of the objects but also represents a lot of texture information. The latter has a lower resolution, blurry edges, and no complex content. The pixel values between the image pairs may have non-linear differences or be totally different. For multi-temporal images, the image content is inconsistent, resulting in the similarity measurement being meaningless. Furthermore, noise also has a great influence on these methods.

Another drawback is that, when using a local window to search for matching, the content in the window should be weakly textured. If the area contained in the window is relatively smooth, there will be a lack of obvious features for matching. This problem also leads to a lot of redundant computations, making the algorithms less efficient.

The area-based methods are not good choices for multi-source remote sensing image registration.

\subsection{Feature-Based Methods}
\label{ssec:featurebased}
The feature-based registration methods detect images' salient structure to extract features as sparse representation, which can greatly reduce the computational complexity. Moreover, the feature-based methods are more robust to the scene difference and noise interference of the images, being relatively efficient.

The commonly used features in image registration include point, line, edge profile, and area features. Among them, compared with other features, the coordinate position of point features can be directly used to calculate the parameters of the transformation model, while other features often need to be transformed into control points. For example, the line feature is represented by the endpoints. Therefore, the point feature is an ideal matching primitive \cite{goshtasby20052}. The feature-based, especially point feature-based algorithm has always been the focus of image registration research and is relatively mature at present. The point feature-based registration algorithm involves the following steps: 1) feature points detection, 2) feature description, 3) feature matching, 4) mismatch removal, 5) parameter estimation and transformation.

Scale-invariant feature transform (SIFT) \cite{lowe1999object,lowe2001local,lowe2004distinctive} is one of the most classic, effective, and commonly used feature point matching methods. It first constructs a Gaussian scale space, supported by scale-space theory \cite{lindeberg1994scale}, and extracts local extreme points in a DoG (Difference of Gaussian) pyramid as the feature points. Then, feature points are described by statistical gradient histograms. SIFT has a good scale and rotation invariance. Many improved algorithms based on SIFT were proposed in recent years. \cite{ke2004pca,mikolajczyk2005performance,bay2006surf,morel2009asift,sedaghat2011uniform,dellinger2014sar,sedaghat2015remote}
Harris et al \cite{harris1988combined} proposed a feature point detection method, which detects the corner structure that changes sharply in multiple directions in an image by calculating corner response according to the gradient. It is an effective and stable feature points detection algorithm, which is widely applied in feature-based image registration.
The oriented FAST and rotated BRIEF (ORB) \cite{rublee2011orb} uses improved accelerated segment test (FAST) \cite{rosten2008faster} and directional binary robust independent elementary features (BRIEF) \cite{calonder2011brief} respectively for feature points detection and description. The algorithm has low time complexity, which is suitable for real-time registration.
Binary robust invariant scalable keypoints (BRISK) \cite{leutenegger2011brisk} and fast retina keypoint (FREAK) \cite{alahi2012freak} are the improved algorithms of BRIEF, which are scale invariant, rotation invariant, and robust to noise. KAZE feature\cite{alcantarilla2012kaze} uses the nonlinear diffusion filtering method to establish the scale-space without blurring the image, different from SIFT.

However, most algorithms are only applicable to single-modal image registration. With the increasing requirements of multi-source image registration, several algorithms for multi-modal image registration were proposed.
Based on SIFT algorithm, Chen et al \cite{alahi2012freak} proposed a partial intensity invariant feature descriptor (PIIFD) for multi-source retinal image registration, which has intensity and rotation invariance but is unable to deal with scale differences.
PSO-SIFT \cite{ma2016remote} introduces a new gradient definition and an enhanced feature matching method for the registration of remote sensing images with intensity differences.
Radiation-variation insensitive feature transform (RIFT) \cite{li2019rift} proposes a maximum index map (MIM) for feature description, which has high intensity invariance, but low rotation invariance, and is not scale invariant.
Most of the research on multimodal image registration is focused on medical images. However, except for some differences in image characteristics and details, the multi-source remote sensing image registration has little difference in essence with medical image registration, so the algorithms also have universality.

\subsection{Deep Learning-Based Methods}
\label{ssec:deeplearning}
In recent years, research has attempted to achieve image registration by using deep learning methods. In general, relevant methods can be divided into two categories \cite{litjens2017survey}: using deep learning networks to estimate the similarity measure of two images to drive iterative optimization, and directly using deep regression networks to predict transformation parameters. Deep learning-based methods first achieved good results in medical image registration. More and more research on remote sensing image registration began to shift its focus to the neural network and deep learning. Although the algorithms based on deep learning are relatively advanced, they often need plenty of annotated samples for model training, in which the model is highly targeted. So the deep learning-based methods cannot be widely applied to all kinds of multi-source data, and are not very mature at present.

%% file: Chap.3.tex
\begin{figure*}[t]
\centering
\centerline{\includegraphics[width=14.5cm]{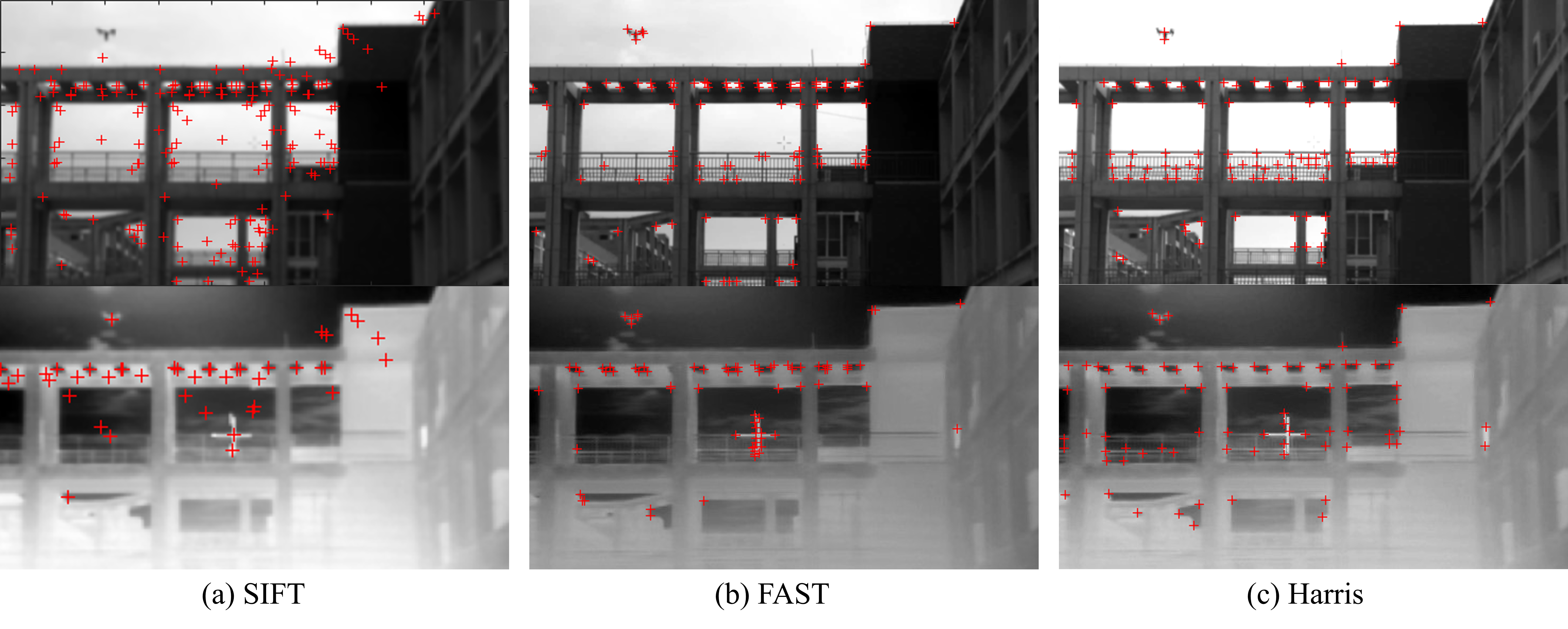}}
\caption{The results of three feature point detection methods in an optical-infrared image pair.}
\label{fig:res}
\end{figure*}

\begin{figure*}[t]
\centering
\centerline{\includegraphics[width=14.5cm]{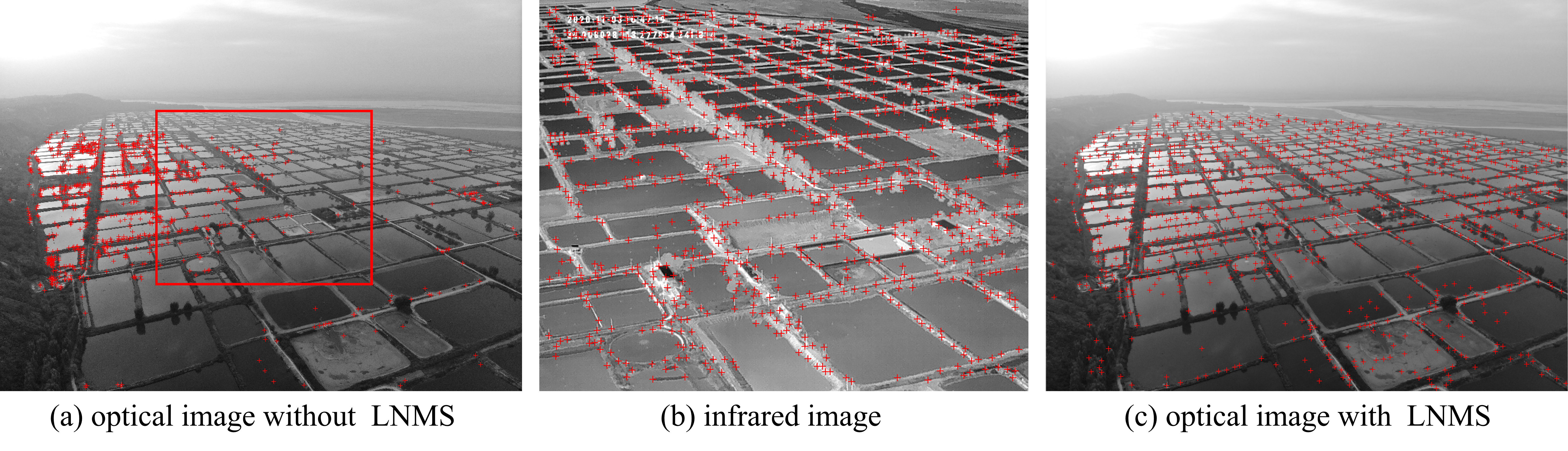}}
\caption{The Harris corner detection results with and without LNMS in an optical-infrared image pair.}
\label{fig:res}
\end{figure*}

The process of the MS-PIIFD registration algorithm is shown in Fig. 1. The input is the image pair to be registered. The preprocessing is carried out first, including simple denoising and data normalization. The images used to obtain features are single-band grayscale images. For multi-band images, such as HSI and MSI, since the proposed algorithm is robust to radiation distortion, it is optional to select one of the bands as the input. Harris corner point detection is carried out on the pre-processed image. Then the scale-space of each image is created, in which a Gaussian pyramid is built. The PIIFD descriptor of each detected Harris corner point is established in the scale space. The feature points are matched according to the descriptors, and then the mismatch removal is carried out. Finally, the spatial correspondence of the matching pairs is used to calculate (estimate) the parameters of the transformation model between the image pair.

\subsection{Feature Point Extraction}
\label{ssec:subhead}
To ensure the accuracy of registration, the feature points should be highly stable and repetitive between the image pair. For multi-source remote sensing images, the difference between scale, blur, intensity, or content details can lead to instability of local features. The feature point detection in SIFT and FAST algorithms is too sensitive to the intensity. Harris corner detection \cite{harris1988combined} focuses on locating obvious 'corner' structure, which is distinct and stable even in multi-model images. Fig. 2 shows the feature point detection results of the three methods in an optical-infrared image pair. By comparison, the feature points obtained by the Harris corner detector are the most highly repetitive. Harris corner detector also has the advantages of strong robustness to local noise and high computation efficiency. Therefore, the Harris corner detector is adopted to extract feature points from multi-source remote sensing images.

Harris corner point is detected by calculating the corner response of each pixel:
\begin{equation}
cornerness = \frac{{{\rm{det}}(\textbf{\emph{M}})}}{{{\rm{tr}}(\textbf{\emph{M}})}}
\end{equation}

\begin{equation}
\textbf{\emph{M}} = 
\begin{bmatrix}
\sum\limits_{(u,v) \in \bf{W}} \bf{I}_x(u,v)^2 & \sum\limits_{(u,v) \in \bf{W}} \bf{I}_x(u,v)\bf{I}_y(u,v)\\
\sum\limits_{(u,v) \in \bf{W}} \bf{I}_x(u,v)\bf{I}_y(u,v) &\sum\limits_{(u,v) \in \bf{W}} \bf{I}_y(u,v)^2
\end{bmatrix}
\end{equation}
where $\rm{det}(\textbf{\emph{M}})$ is the determinant of $\textbf{\emph{M}}$, $\rm{tr}(\textbf{\emph{M}})$ is the trace of $\textbf{\emph{M}}$, ${{{\bf{I}}_x}(u,v)}$ and ${{{\bf{I}}_y}(u,v)}$ are the gradient of the image in the $x$ and $y$ directions, ${\bf{W}}$ is a gaussian window.

One key point worth noting is that multi-source remote sensing images may have serious intensity distortion or scale difference, which results in the nonuniform distribution and low repeatability of feature points in the image pair. This problem is visually illustrated in Fig. 3, which is the Harris corner detection result of an optical-infrared image pair with a large scale difference. The result of directly filtering with a threshold of the optical image is shown in Fig. 3(a), in which most feature points are concentrated in the left part of the image. However, the approximate position of the infrared image in the visible image is indicated by the red box, and only a small number of feature points are distributed in it, which are very nonuniform. The proposed algorithm adopts local non-maximum suppression (LNMS) to solve this problem. From Fig. 3(c), it can be seen that, with LNMS, the feature points are far more uniformly distributed, and the repeatability is higher.

\subsection{Multi-scale PIIFD}
\label{ssec:subhead}
PIIFD \cite{chen2010partial} is an improved algorithm based on SIFT \cite{lowe2004distinctive} to solve the problem of the full inverse (light and dark flipping) and nonlinear intensity difference in multi-source retinal images. Similar problems are evident in multi-source remote sensing images, which can be seen through the image pairs in Fig. 2 and Fig. 3. So in this paper, the PIIFD algorithm is used to describe the feature points.

In the original PIIFD \cite{chen2010partial}, at first, each feature point is given a main orientation (0 $\sim\pi$), by calculating the local second-order gradient, which ensures rotation invariance. Then the neighborhood of each feature point, the size of which is fixed at $40*40$ pixels, is used for description. Taking the main orientation as the datum (0$^{\circ}$), the gradient within the neighborhood is counted. Finally, a descriptor vector of length $128$ is generated for feature point description. However, for multi-source remote sensing images with different scales, the information in the local neighborhood can have a large difference, which will lead to different descriptors of the same point.

\begin{figure}[t]
\centering
\centerline{\includegraphics[width=6.5cm]{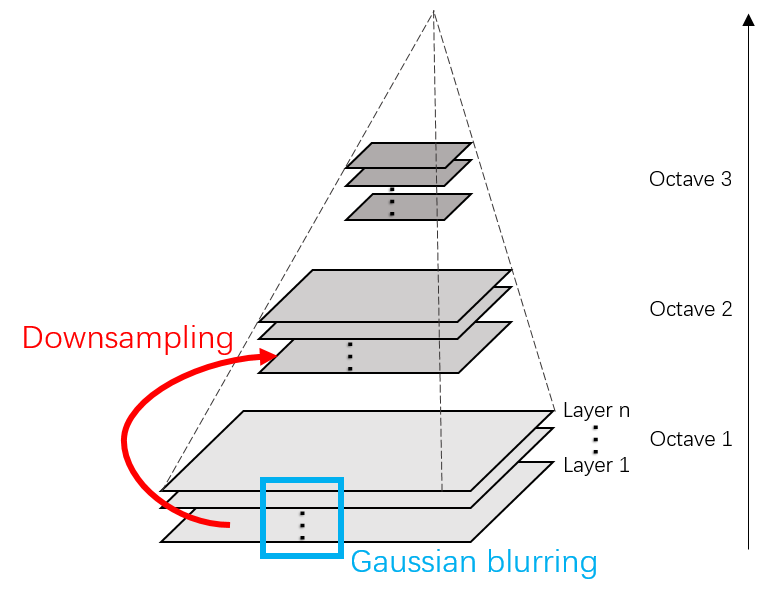}}
\caption{Structure of the scale space (gauss pyramid) used in MS-PIIFD.}
\label{fig:res}
\end{figure}

In order to realize multi-scale PIIFD, local information of feature points needs to be extracted in the scale space of the images. The method of building image's Gaussian pyramids based on the scale-space theory \cite{lindeberg1994scale} is adopted. The schematic diagram of establishing Gaussian pyramid of image in the proposed algorithm is shown in Fig. 4. The original image is sampled down step by step to obtain a series of images with different scales, that is, the first layer in each octave. Then a series of Gaussian blurs are performed on each octave:
\begin{equation}
{\bf{L}} = {\bf{G}} * {\bf{I}}
\end{equation}
\begin{equation}
{\bf{G}} = \frac{1}{{\sqrt {2\pi {\sigma ^2}} }}{e^{\frac{{ - ({x^2} + {y^2})}}{{2{\sigma ^2}}}}}
\end{equation}
where ${\bf{I}}$ is the original image, ${\bf{G}}$ is a Gaussian kernel with standard deviation of $\sigma$, and ${\bf{L}}$ is the Gaussian blur image.

After the scale-space of the images is established, for each Harris corner point, the multi-scale PIIFD descriptor is calculated by obtaining the neighborhood information at the corresponding location in the scale space. The algorithm flow is shown in Algorithm 1, where ${\bf{G}}_{\bf{I}}(x,y,O,L)$ is the Gaussian scale space of image $\bf{I}$ established through down-sampling and Gaussian blur, $O$ is the number of octave with different scale, $L$ is the number of layers with different blurring degree. The algorithm outputs the feature point descriptor set $D_{P}$, which contains 128-dimensional descriptors for each feature point at each scale. The feature described in the scale-space can reduce the influence of scale difference of multi-source images and obtain the best matching result of feature points.

\begin{algorithm}[htp]
\caption{Proposed multi-scale PIIFD for registration}
\hspace*{0.02in} {\bf Input:}
feature point set $P_{\bf{I}}$, image scale space \\\hspace*{0.47in}${\bf{G}}_{\bf{I}}(x,y,O,L)$, PIIFD window scale $S$\\
\hspace*{0.02in} {\bf Output:}
feature descriptor set $D^{P}$
\begin{algorithmic}[1]
\State In each layer of ${\bf{G}}_{\bf{I}}(x,y,O,L)$:
\State \hspace*{0.2in}Calculate the second-order gradient of the image
\State \hspace*{0.2in}Calculate the first-order gradient of the image
\State \hspace*{0.2in}For each feature point $P$ in $P_{\bf{I}}$:
\State \hspace*{0.4in}Calculate the corresponding position
\State \hspace*{0.4in}Calculate the main orientation using second-order gradient
\State \hspace*{0.4in}Taking the main orientation as the reference direction ($0^{\circ}$), establish a square window with size (side length) of S, and divide it into 16 small squares
\State \hspace*{0.4in}Statistics the first-order gradient direction within each small square to obtain feature descriptor $D(P,O,L)$
\State Output feature descriptor set $D_{P_{\bf{I}}}$
\end{algorithmic}
\end{algorithm}

\subsection{Matching and Transformation}
\label{ssec:subhead}
After all the descriptors of each feature point are created, the best-bin-first (BBF) \cite{beis1997shape} matching method is adopted to obtain the initial matches of the feature points, when the cosine similarity is used as the measurement. For multi-source images, there is likely to be a large number of mismatches in initial matches, which causes common outlier removal algorithms to fail, such as RANSAC (random sample consensus). In the proposed algorithm, the consistency of the main orientation and spatial distribution of the feature points \cite{chen2010partial} are used for mismatch removal. This is because although there may be spatial distortion in the multi-source remote sensing image pairs, there will not be too much deformation on the whole. Then, assuming that there is only overall rotation or scaling between the image pair. So the angle between the main orientation of each pair of correctly matched feature points should be consistent, and their spatial relationship with other correctly matched points in the two images should also be consistent. Therefore, this method is suitable for multi-source remote sensing images.

Finally, three transformation models, similarity, affine, and projective transformation, are selected for parameter estimation and image transformation to deal with the different spatial distortion that may exist between the image pairs, where the least square method is adopted to estimate parameters.

%% file: Chap.4.tex
\begin{figure*}[t]
\centering
\centerline{\includegraphics[width=17.5cm]{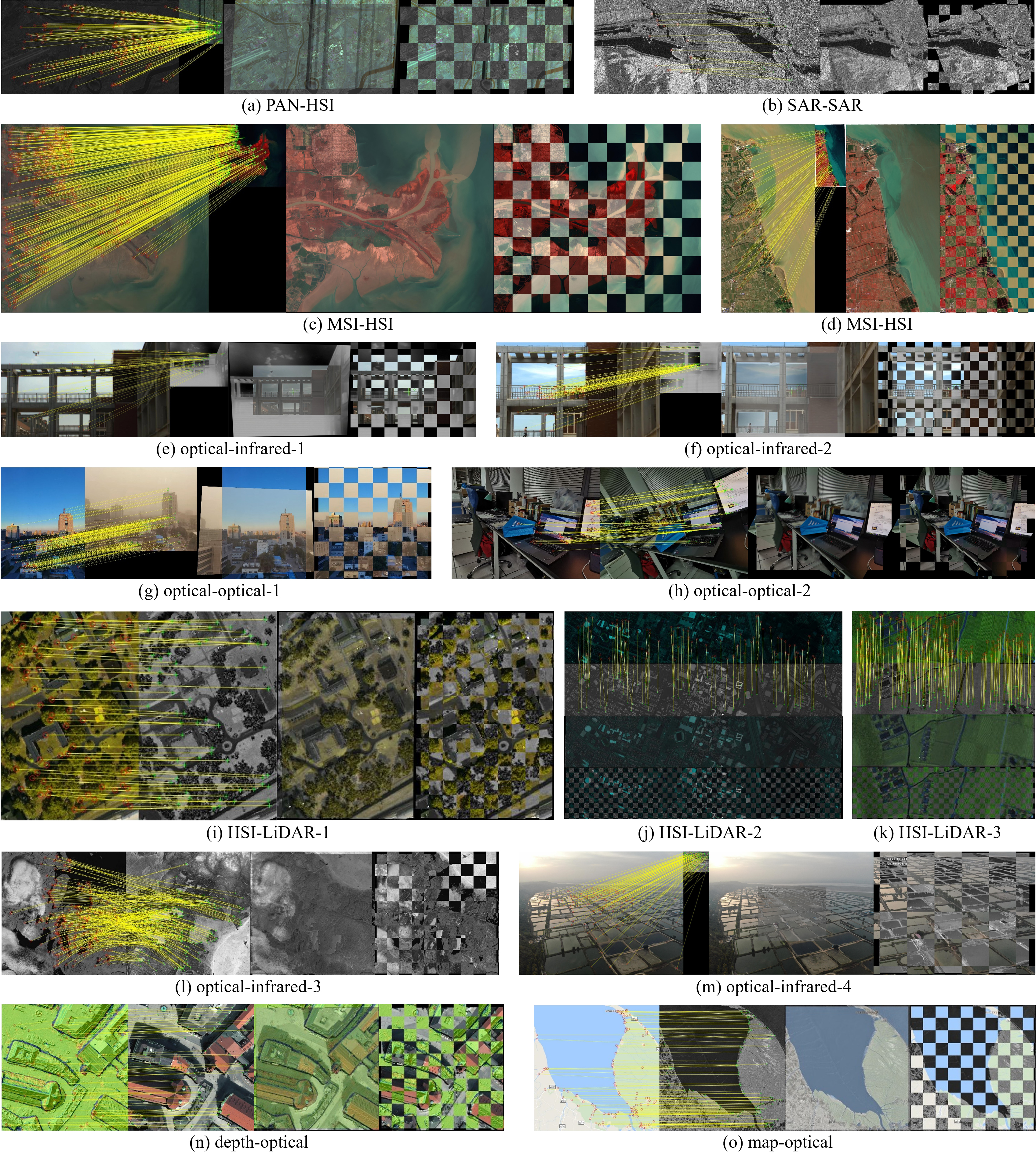}}
\caption{Registration results of the proposed MS-Harris-PIIFD in three forms, including matching results (left), fusion form (middle), and checkerboard form (right).}
\label{fig:res}
\end{figure*}

\renewcommand{\arraystretch}{1.5}
\begin{table*}[!htb]
  \centering
  \fontsize{6.5}{8}\selectfont
  \caption{Number of correct matching pairs comparison by five registration methods.}
  \label{tab:performance_comparison}
    \begin{tabular}{|c|c|c|c|c|c|c|c|c|c|c|c|c|c|c|c|}
    \hline
    \multirow{2}{*}{Method}&
    \multicolumn{15}{c|}{Scene}\cr\cline{2-16}
    &a&b&c&d&e&f&g&h&i&j&k&l&m&n&o\cr
    \hline
    \hline
    SIFT \cite{lowe2004distinctive}&22&14&112&5&0&0&12&{\bf 108}&6&9&10&0&16&0&0\cr\hline
    PIIFD \cite{chen2010partial} &0&0&0&0&0&0&48&0&0&0&12&180&0&18&17\cr\hline
    PSO-SIFT \cite{ma2016remote} &0&{\bf 21}&58&0&6&18&22&107&16&21&8&127&0&8&0\cr\hline
    RIFT$^{+}$ \cite{li2019rift} &0&7&7&3&2&2&{\bf 190}&0&{\bf 158}&{\bf 504}&76&0&0&{\bf 410}&{\bf 214}\cr\hline
    Proposed &{\bf 181}&16&{\bf 383}&{\bf 90}&{\bf 63}&{\bf 114}&175&76&55&271&{\bf 155}&{\bf 293}&{\bf 112}&82&113\cr\hline
    \end{tabular}
\end{table*}

To validate the effectiveness of the proposed MS-PIIFD, a distinct set of multi-source image data are used to test the robustness and universality of the proposed algorithm, and four state-of-the-art multi-source registration algorithms including SIFT \cite{lowe2004distinctive}, PIIFD \cite{chen2010partial}, PSO-SIFT \cite{ma2016remote} and RIFT$^{+}$ \cite{li2019rift} are selected for comparison. Except for the feature point detection and description method, all processing is the same, including image normalization, noise reduction, feature point matching, and mismatch removal, etc. The experiments are all implemented using MATLAB2021a on a platform with Intel Core I7-8700 CPU.

\subsection{Experimental Data}
\label{ssec:data}
The data used in experiments are multi-source image pairs of 15 scenes, labeled a$\sim$o, in which there are both multi-sensor and multi-temporal images, including HSI, MSI, LiDAR-derived DSM (digital surface model), SAR, optical images, infrared images, depth map, and artificially produced rasterized map. These selected experimental data basically contain all the problems in multi-source image registration, including large resolution differences, serious radiance distortions, spatial distortion, rotation, content details differences, and image noise. To further test the universality of the proposed algorithm, not only satellite data but also UAV data and ground scene data are selected.

\subsection{Parameter Setting}
The proposed MS-PIIFD algorithm contains four main parameters that can greatly affect the registration, which are the size of the LNMS window and the threshold of the number of feature points in the Harris corner detector, and the total number of octaves and layers of the Gaussian pyramid.

In Harris corner points detection, since the positional relationship and scale difference of the image pairs are unknown in advance, we hope that the feature points are distributed as uniformly as possible in the image pair. So, the ratio of the size of the LNMS window depends on the ratio of the pixel number of the two images:
\begin{equation}
ratio = \sqrt {\frac{{M \times N}}{{m \times {\rm{n}}}}}
\end{equation}
where $M,N$ and $m,n$ are the length and width of the two images. In addition, the minimum unit size in the established PIIFD descriptor is $10\times10$. Thus, the minimum size of the LNMS window is set to $10\times10$, to prevent poor separability of the descriptors due to the feature points being excessively close. High repetition of feature points is required to provide sufficient matching candidates, so it is important to provide sufficient points to be matched. But too many feature points not only increase the computational burden but also bring a lot of outliers. Through experiments, the threshold of the Harris corner points number is set between 500 and 1000. The specific number is determined according to the size of the image. In this experiment, the maximum number of points was set at 1000.

To ensure the efficiency of registration, it is necessary to set the suitable total number of octaves and layers, when too large scale space will increases the computational burden a lot. Through experiments, the total numbers of octaves and layers are set to 3 and 4, which can basically handle most multi-source image pairs.

\subsection{Visual Results}
\label{ssec:visual}
The visual registration results of the proposed algorithm are shown in Fig. 3. The results of each scene are respectively displayed in three forms: matching results of feature points, fusion form, and checkerboard form of the image pairs. Although the characteristics of scene contents of multi-source images are not consistent, the alignment of the images can be evaluated by roughly comparing the coincidence of objects' edges.

By artificial comparison, there is no obvious deviation or 'virtual shadow' in the images, except for scene f. This is because the imaging quality of the infrared image is low, and the image has a large spatial distortion, especially at the edge of the image. Simply using similarity, affine, or projection transformations will not align the image in all parts. In general, some functional interpolation and fitting methods will be applied in this case to achieve a better alignment effect, but it is not the focus of this paper, and further research will pay attention to these issues. It can be found that the spatial difference of images is within 1$\sim$2 pixels. Apart from the edge positions of scene f, there is no obvious registration failure in any of the scenes, and the registration algorithm is effective.

\subsection{Numerical Results}
\label{ssec:numerical}
The registration accuracy is usually measured by root mean squared error (RMSE) and other similar methods, through comparing the similarity of the image pair after registration. However, since the intensity of multi-source, especially multi-modal images, is different, the methods of measuring similarity do not have physical significance. Therefore, the registration effect is evaluated by directly comparing the number of correct matches of the feature points, because the number of correct control points is positively correlated with the accuracy of the transformation model. When the number of matched feature points is less than 3 pairs, the parameters of the spatial transformation model cannot be solved, and the registration is considered to be failed.

The numbers of correct matching pairs of the five registration methods are shown in Table 1.
In scene b, the proposed algorithm gets relatively few matches, and 5 less than PSO-SIFT. Through analysis, it is because that the texture information of the SAR image is relatively complex and chaotic, which reduces the performance of the proposed algorithm, but still obtains enough correct matches. In the scene g, h, i, j, n, o, the proposed algorithm does not get the most matches, and in contrast, the Rift+ is doing better. It is because there is basically no scale difference between the selected images of these scenes. In particular, the scene i, j, n, o are image pairs that have been artificially modified to the same scale. However, the proposed MS-PIIFD still obtains a considerable number of correct matches. Among all the competitive methods, the proposed registration algorithm is the only one that does not fail in any scene and has significant advantages in most scenes.

%% file: Chap.5.tex
In this paper, an effective image registration algorithm based on MS-PIIFD was proposed for multi-source remote sensing image registration. Through experiments and analysis, the algorithm can well achieve the registration of multi-source remote sensing images and performs well in practical work. The image registration error can reach within 1$\sim$2 pixels, and the objects in the scene can be aligned accurately. Through tests of various scene data registration, it is concluded that the algorithm has good robustness, stability, and universality, and has strong practical application value.